\documentclass[prb,longbibliography,twocolumn,superscriptaddress,amsmath,amssymb,verbatim]{revtex4-1}

\usepackage{graphicx}
\usepackage{lmodern}
\usepackage{epstopdf}
\usepackage[colorlinks=true,citecolor=blue,linkcolor=blue,urlcolor=blue,bookmarks=true]{hyperref}
\usepackage{xcolor}
\definecolor{green}{rgb}{0.1, 0.6, 0.3}

\begin{document}

\preprint{APS/123-QED}

\title{Magnetic inhomogeneity in the copper pseudochalcogenide CuNCN}

\author{A. Zorko}
\email{andrej.zorko@ijs.si}
\affiliation{Jo\v{z}ef Stefan Institute, Jamova c.~39, SI-1000 Ljubljana, Slovenia}
\author{P. Jegli\v c}
\affiliation{Jo\v{z}ef Stefan Institute, Jamova c.~39, SI-1000 Ljubljana, Slovenia}
\author{M. Pregelj}
\affiliation{Jo\v{z}ef Stefan Institute, Jamova c.~39, SI-1000 Ljubljana, Slovenia}
\author{D. Ar\v con}
\affiliation{Jo\v{z}ef Stefan Institute, Jamova c.~39, SI-1000 Ljubljana, Slovenia}
\affiliation{Faculty of Mathematics and Physics, University of Ljubljana, Jadranska 19, SI-1000 Ljubljana, Slovenia}
\author{H. Luetkens}
\affiliation{Laboratory for Muon Spin Spectroscopy, Paul Scherrer Institute, CH-5232 Villigen PSI, Switzerland}
\author{A. L. Tchougr\'eeff}
\affiliation{Institute of Inorganic Chemistry, RWTH Aachen University, D-52056 Aachen, Germany}
\affiliation{Moscow Center for Continuous Mathematical Education, Bol. Vlasevskyi 11, 119002 Moscow, Russia}
\affiliation{A.~N.~Frumkin Institute of Physical Chemistry and Electrochemistry of Russian Academy of Science, Moscow, Russia}  
\author{R. Dronskowski}
\affiliation{Institute of Inorganic Chemistry, RWTH Aachen University, D-52056 Aachen, Germany}

\date{\today}

\begin{abstract}
Copper carbodiimide, CuNCN, is a geometrically frustrated nitrogen-based analog of cupric oxide,  whose magnetism remains ambiguous.
 Here, we employ a combination of local-probe techniques, including $^{63,\, 65}$Cu nuclear quadrupole resonance, $^{13}$C nuclear magnetic resonance and muon spin rotation to show that the magnetic ground state of the Cu$^{2+}$ ($S=1/2$) spins is frozen and disordered. 
 Moreover, these complementary experiments unequivocally establish the onset of an intrinsically inhomogeneous magnetic state at $T_h=80$~K. 
Below $T_h$, the low-temperature frozen component coexists with the remnant high-temperature dynamical component down to $T_l = 20$~K, where the latter finally ceases to exist.
Based on a scaling of internal magnetic fields of both components we conclude that the two components coexist on a microscopic level.  
\end{abstract}

\maketitle

\section{Introduction}
Geometrical frustration, where lattice geometry prevents conventional magnetic ordering, is one of the major promoters of exotic phenomena in condensed matter.
In geometrically frustrated antiferromagnets, perplexing magnetic states, such as spin glasses and spin liquids, with unconventional excitations are regularly encountered.\cite{lacroix2011introduction, balents2010spin}
Furthermore, even in uniform spin systems, competition among various nearly degenerate phases can lead to inhomogeneous magnetic states on a microscopic scale.\cite{schmalian2000stripe,mu2002stripe, kamiya2012formation}
Various intriguing experimental cases that fall in this category have indeed been reported. 
The first type of examples are systems in which a fraction of all spins remains dynamical while the rest either order\cite{stewart2004phase, zheng2006coexisting, ling2017striped} or form valence bonds.\cite{nakajima2012microscopic} 
Further examples are realizations of multiple magnetic orders that locally compete,\cite{zorko2014frustration, zorko2015magnetic, nilsen2015complex} and a nanoscale modulation of magnetic order.\cite{pregelj2015spin, pregelj2016exchange}
\begin{figure}[b]
\includegraphics[trim = 0mm 0mm 0mm 0mm, clip, width=1\linewidth]{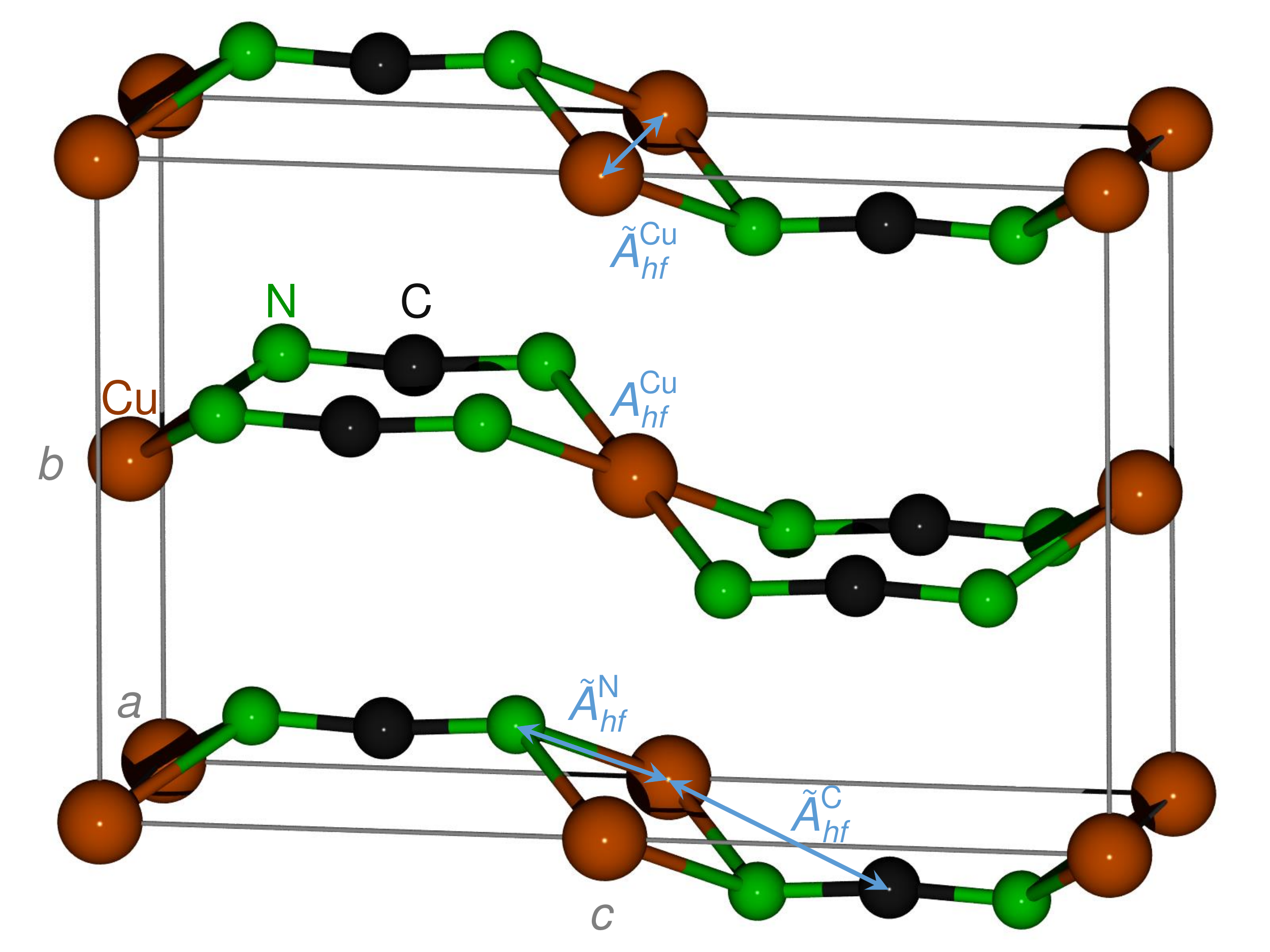}
\caption{The unit cell of CuNCN. The on-site hyperfine interaction of the $^{63,65}$Cu nuclei with the Cu$^{2+}$ electronic moment is denoted by $A_{hf}^{\rm Cu}$, while the transferred hyperfine interactions of various nuclei $n$ with the Cu$^{2+}$ electronic moment are indicated by arrows and labeled as $\tilde{A}_{hf}^{n}$.}
\label{fig0}
\end{figure}


The nitrogen-containing analog of cupric oxide, CuNCN,\cite{liu2005novel} is a potential contender for such a class of antiferromagnets.
Its unit cell (Fig.~\ref{fig0}) has orthorhombic symmetry and corresponds to the $Cmcm$ space group.  
It possesses a gap at the Fermi surface\cite{liu2008characterization} and magnetic exchange coupling between $S=1/2$ spins localized on the Cu$^{2+}$ sites is predicted to be of the order of several hundreds of kelvins.\cite{tsirlin2010uniform, tchougreeff2013low}
Nevertheless, various experiments initially suggested that this compound lacks classical long-range magnetic ordering.\cite{liu2008characterization, xiang2009theoretical}  
Strong exchange coupling explains a very small magnetic susceptibility, which
exhibits an anomaly at $T_h \simeq 80$~K that remains controversial.
A broad feature was observed around this temperature also in magnetic heat capacity, where another feature was witnessed around $T_l \sim 20$~K.\cite{tchougreeff2017atomic} 
The  $T_h$ anomaly was theoretically ascribed either to long-range magnetic ordering\cite{tsirlin2010uniform,tsirlin2012hidden} or to a spin-liquid instability.\cite{tchougreeff2013low, tchougreeff2014mean}
Spin-liquid instabilities, where systems cross from one spin-liquid state to another, have been experimentally detected in frustrated antiferromagnets on a few occasions. \cite{itou2010instability,gomilsek2016instabilities,gomilsek2017field,klanjsek2017high}
In CuNCN, the spin-liquid instability scenario assumes a transition from a gapless high-temperature spin liquid into a pseudo-gapped low-temperature spin liquid at $T_h$.\cite{tchougreeff2013low, tchougreeff2014mean} 
This scenario can indeed offer an explanation for the observed susceptibility anomaly,\cite{zorko2011unconventional} as well as for  accompanying structural changes\cite{tchougreeeff2012structural, jacobs2013high} and for the unusual temperature dependence of the magnetic heat capacity revealing a relatively small amount of entropy released up to 150~K.\cite{tchougreeff2017atomic}
On the other hand, the alternative scenario of long-range magnetic order is based on the detection of frozen local magnetic fields in muon spin relaxation ($\mu$SR) experiments\cite{zorko2011unconventional, tsirlin2012hidden} and on strong quantum spin fluctuations that are predicted to reduce the ordered part of the magnetic moments beyond the sensitivity of most experimental techniques.\cite{tsirlin2012hidden} 
However, the $\mu$SR experiments have further revealed that the width of the static-local-field distribution is very large, being of the same size as the average field magnitude even at temperatures as low as 63~mK.\cite{zorko2011unconventional}
This is clearly not a characteristic of an antiferromagnetic long-range order.

Quite surprisingly, in CuNCN the frozen state, as detected by $\mu$SR, seems to be fully established over the entire sample only below $T_l$, while in the broad temperature range $T_h \gtrsim T \gtrsim T_l$ the $\mu$SR signal attributed to the remnant dynamical component progressively disappears with decreasing temperature (Fig.~\ref{fig1}b).\cite{zorko2011unconventional}
A similar conclusion of two coexisting magnetic components was drawn from $^{14}$N nuclear magnetic resonance (NMR) experiments, which also disclosed a broad distribution of local environments below $T_h$.\cite{zorko2011unconventional}
Yet, both magnetic components have not been so far unambiguously simultaneously detected and evaluated in a single experiment.  
Furthermore, the assessment of the intrinsic nature of such intricate magnetism has not been provided either.
In principle, static magnetism and the coexistence of two magnetic components could be triggered by external perturbations, e.g., by muons perturbing the local environment in the $\mu$SR experiments and by strong applied magnetic fields in the NMR experiments,\cite{zorko2011unconventional} if the dynamical disordered state was unstable.

All these open questions about the magnetic inhomogeneity and the true nature of the magnetic ground state in CuNCN thus call for complementary experimental approaches that will clarify previous results and confront the existing theoretical proposals.
Therefore, we have performed nuclear quadrupole resonance (NQR) experiment, as well as additional NMR and $\mu$SR experiments.
The NQR investigation, which is performed in zero applied field, eliminates possible effects of both the applied magnetic field and external probes on magnetism of CuNCN. 
The $^{13}$C NMR has been chosen because the hyperfine coupling (Fig.~\ref{fig0}) of the $^{13}$C nuclei to the electronic magnetism is much smaller than that of $^{14}$N nuclei, thus allowing for simultaneous detection of both static and dynamical magnetic components below $T_h$. 
The $^{13}$C NMR thus complements the previous $^{14}$N NMR measurements that could detect only the dynamical magnetic component below $T_h$, while the static component was inaccessible.\cite{zorko2011unconventional}
Finally, simultaneous detection of the two components is achieved also by a complementary $\mu$SR experiment in a strong transverse magnetic field (TF) that, in contrast to previous $\mu$SR experiments,\cite{zorko2011unconventional, tsirlin2012hidden} by far exceeds the internal static fields. 
The new results reveal intrinsic coexistence of the two fundamentally different magnetic components, i.e., a dynamical and a static one, which in CuNCN compete on a microscopic scale in the broad temperature range between $T_h=80$~K and $T_l=20$~K.

\section{Experimental Details}
All experiments were performed on the same high-quality batch of CuNCN powders, as used in our previous local-probe magnetic investigations.\cite{zorko2011unconventional}

The NQR spectra of the $^{63}$Cu nuclei with the natural abundance of 69\% and the $^{65}$Cu nuclei with the natural abundance of 31\% were recorded on a custom-built spectrometer using a solid-echo pulse sequence $\beta-\tau-\beta-\tau-{\rm echo}$ with the $\pi/2$-pulse length $\beta=4.8$~$\mu$s and the interpulse delay $\tau=20$~$\mu$s, by sweeping the frequency in $\Delta\nu = 50$~kHz steps.
The nuclear spin-spin relaxation time $T_2$ was measured with the same pulse sequence on the $^{63}$Cu NQR peak by varying the delay $\tau$.
The gyromagnetic ratios of the two isotopes are $^{63}\gamma = 2\pi\cdot 11.28$~MHz/T and $^{65}\gamma = 2\pi\cdot 12.09$~MHz/T.

The $^{13}$C NMR spectra were recorded in a magnetic field of 9.4~T with parameters $\beta=9$~$\mu$s, $\tau=30$~$\mu$s and $\Delta\nu = 30$~kHz.
The Larmor frequency in this case, given by the $^{13}$C nuclear gyromagnetic ratio $^{13}\gamma = 2\pi\cdot 10.71$~MHz/T, was $\nu_0=100.571$~MHz. 
An inversion recovery method was used for $^{13}$C spin-lattice relaxation ($T_1$) measurements, where the spectra were integrated within an 80~kHz window centered at positions $\nu_A=100.600$~MHz or $\nu_B=100.450$~MHz.

The $\mu$SR experiment was performed on the General Purpose Surface-Muon Instrument (GPS) at the Paul Scherrer Institute, Switzerland.
It was conducted in a transverse-muon-polarization mode in the magnetic field of 0.52~T applied along the muon beam direction and perpendicular to the up-down positron detectors.
The field yielded muon-polarization precession at the Larmor frequency of $70.46$~MHz, given by the muon gyromagnetic ratio $\gamma_\mu =2\pi\cdot 135.5$~MHz/T. 
The angle between the initial polarization and the beam direction was about $45^{\circ}$, which resulted in the full positron-detector (muon) asymmetry of $A_0=0.206$.

The muons, being spin-1/2 entities, can serve as efficient local probes of magnetism,\cite{yaouanc2011muon} as they are initially almost $100\%$ polarized in a $\mu$SR experiment. 
In a local field their polarization starts precessing, which is effectively measured by the direction of emitted positrons, arising from muon decays.
The resulting muon asymmetry, measured by oppositely facing positron detectors with respect to the sample, is linearly proportional to the polarization.

\section{Results}

\subsection{$^{63,\,65}$Cu Nuclear Quadrupole Resonance}
All previous magnetic studies of CuNCN relied on applying a finite magnetic field.\cite{liu2008characterization, zorko2011unconventional, tsirlin2012hidden}
In the case of a competition among different nearly-degenerate candidate ground states, as it may occur in geometrically frustrated systems,\cite{lacroix2011introduction} the applied field can give preference to a particular state.
This was suggested previously as a possible explanation of the unusual magnetism of CuNCN below $T_h$.\cite{zorko2011unconventional}
The current NQR experiment performed in zero applied field would eliminate the field-induced effect if it was really present.

\begin{figure}[t]
\includegraphics[trim = 0mm 0mm 0mm 0mm, clip, width=1\linewidth]{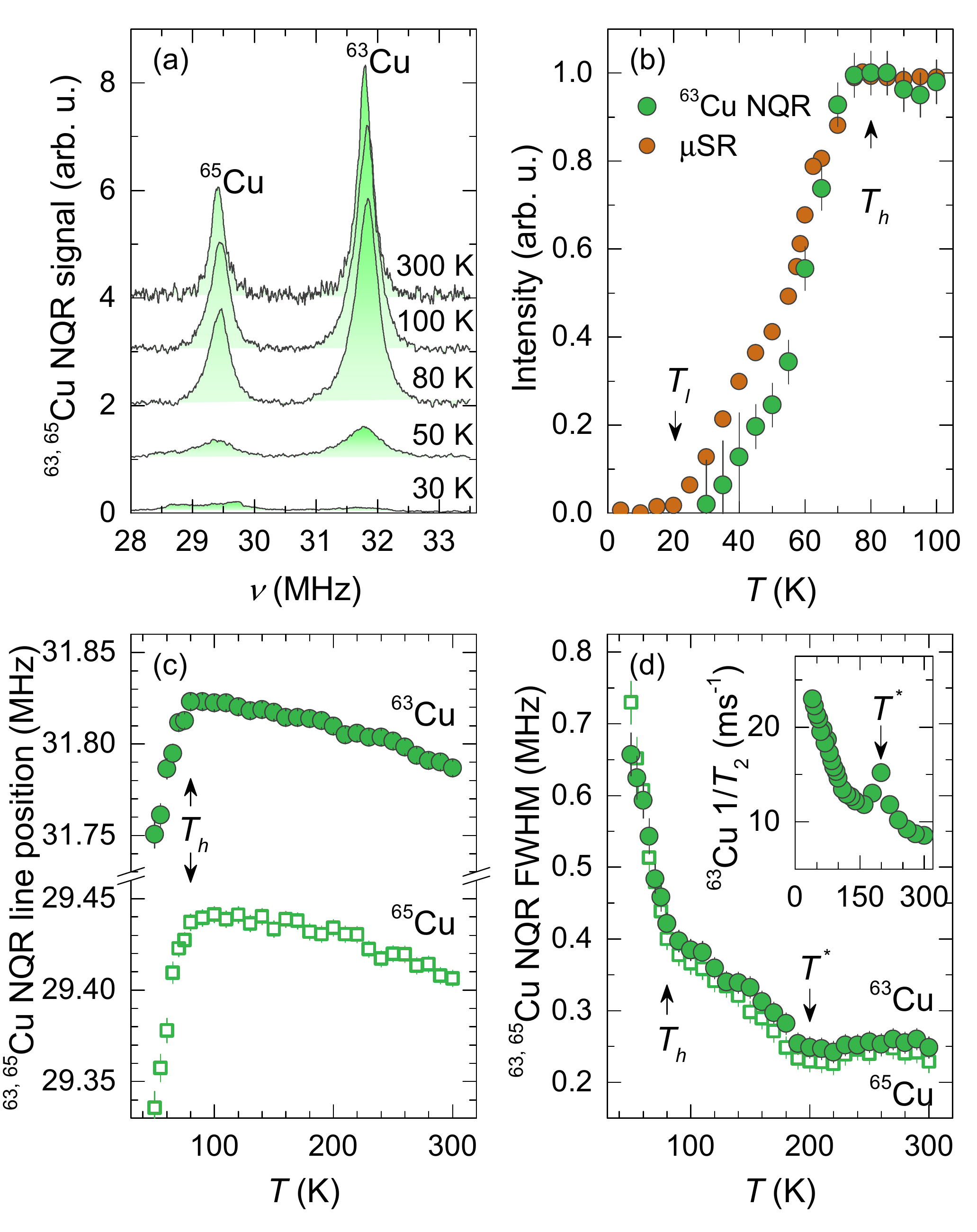}
\caption{(a) The $^{63,\,65}$Cu NQR spectra of CuNCN at a few selected temperatures.
The spectra are shifted vertically for clarity and their intensity is normalized by the Boltzmann population factor $1/T$ and corrected for the spin-spin relaxation time $T_2$ [inset in (d)]. 
(b) The temperature dependence of the corrected $^{63}$Cu NQR line intensity compared to the unfrozen part of the $\mu$SR signal (reproduced from Ref.~\onlinecite{zorko2011unconventional}). 
The temperature dependence of the $^{63,\,65}$Cu NQR (c) line position and (d) line width.
The inset displays the temperature dependence of the $^{63}$Cu NQR spin-spin relaxation rate.
The arrows indicate special temperatures $T_h$, $T_l$ and $T^*$.}
\label{fig1}
\end{figure}

In CuNCN, there exists a single Cu crystallographic site.
Since both Cu isotopes, $^{63}$Cu and $^{65}$Cu, possess nuclear spin $I=3/2$, each isotope gives a single NQR line (corresponding to the transitions $I_z=\pm 3/2 \leftrightarrow \pm 1/2$) when no static magnetic field is present in a material.\cite{abragam1961principles}
Consequently, the $^{63,\, 65}$Cu NQR spectrum of CuNCN consists of two spectral lines (Fig.~\ref{fig1}a). 
The $^{63}$Cu line is about twice as intense due to the larger natural abundance of this isotope.
The position of each line is given by the nuclear quadrupolar moment $Q$ of each isotope and the common electric-field-gradient (EFG) tensor of the crystal at the Cu site.
As $^{63}Q/^{65}Q=1.08$, the $^{63}$Cu spectral line is positioned at a higher frequency compared to the $^{65}$Cu line.
Accordingly, we find the $^{63}$Cu line at 31.79~MHz and the $^{65}$Cu line at 29.41~MHz at room temperature.
The full-width-at-half-maximum (FWHM) of the $^{63}$Cu NQR line $^{63}\delta=250(10)$~kHz is larger than the $^{65}$Cu FWHM $^{65}\delta=230(10)$~kHz by the same factor $^{63}Q/^{65}Q$, demonstrating that the NQR line width in CuNCN is due to a distribution of EFGs.
The measured widths of several hundred kHz must have an entirely static origin, since homogeneous broadening ($1/T_2$) due to spin-spin relaxation is of the order of only 10~kHz (inset in Fig.~\ref{fig1}d). 

The $^{63,\,65}$Cu NQR line position (Fig.~\ref{fig1}c) and line width (Fig.~\ref{fig1}d) exhibit rather pronounced temperature dependences.
The line position, which is sensitive to the changes in the crystal lattice through the changes in the EFG tensor, exhibits an unusual maximum at $T_h=80$~K (Fig.~\ref{fig1}c).
This is in line with a nonmonotonic behavior of lattice parameters that has been recently observed in synchrotron\cite{tchougreeeff2012structural} and neutron-diffraction experiments. \cite{jacobs2013high}
Moreover, also the line width exhibits an anomaly at $T_h$, where it starts increasing rapidly on decreasing temperature (Fig.~\ref{fig1}d), reminiscent of the $^{14}$N NMR case.\cite{zorko2011unconventional}
Although the $^{63}$Cu NQR line width is larger than the $^{65}$Cu line width at high temperatures, the situation is reversed below $T_h$.
At 50~K we find the ratio $^{63}\delta/^{65}\delta=0.90(3)$, which corresponds well to the gyromagnetic-factors ratio $^{63}\gamma/^{65}\gamma=0.93$.
Therefore, in addition to the EFG NQR broadening mechanism, there is a second broadening mechanism of magnetic origin that prevails below $T_h$. 
This demonstrates the growth of static magnetic fields at copper sites below $T_h$ on the NQR time scale.

The magnitude of static magnetic fields below $T_h$ can be estimated from the magnetic broadening.
Since the NQR line is a convolution of two broadening mechanisms, the square of the full width is a sum of squared individual contributions.
For the $^{65}$Cu line at 50~K the full line width of $730(30)$~kHz and the quadrupolar contribution of the order of $250(10)$~kHz yield the magnetic line width of $\delta^{65}=690(30)$~kHz, which corresponds to a magnetic field of $\pi\, \delta^{65}/^{65}\gamma = 30(1)$~mT.
This should be compared to a typical on-site field of 25~T felt by the nucleus due to the hyperfine coupling with the fully polarized, on-site Cu$^{2+}$ ($S=1/2$) electronic spin.\cite{abragam19702electron}
The observed, relatively small broadening of the NQR spectra is thus clear evidence that the magnetic state of CuNCN, as detected by $^{63,\,65}$Cu NQR, remains predominantly dynamical even below $T_h$.

Next, we highlight another intriguing property of the low-temperature NQR spectra, namely, a loss of the signal, i.e., a spectral wipe-out, that is observed to occur progressively below $T_h$ (Fig.~\ref{fig1}a).
Namely, the NQR spectral intensity, when corrected for the $T_2$ relaxation effect (inset in Fig.~\ref{fig1}d) and the Boltzman population factor $1/T$, shows a sudden drop below $T_h$, yet it vanishes completely only around $T_l$  (Fig.~\ref{fig1}b).
This atypical behavior resembles the slow disappearance of the dynamical signal in the $\mu$SR experiment\cite{zorko2011unconventional} (Fig.~\ref{fig1}b),
signaling that this is an intrinsic effect and not an effect induced by external perturbations, such as the implanted muons or the applied magnetic field.

We note that the disappearance of the NQR signal speaks against two possible scenarios related to spin liquids. Firstly, in a spin liquid, an enhancement of nuclear relaxation can lead to a spectral wipe-out.\cite{zorko2008easy} 
Yet, in CuNCN, there is no drastic enhancement of the $T_2$ relaxation, which is always faster than the $T_1$ relaxation.
Secondly, if a transition from a gapless to a gapped spin liquid were taking place at $T_h$,\cite{tchougreeff2013low,tchougreeff2014mean} the opening of a spin gap would reduce spin excitations and a full-intensity NQR line would remain.  
Hence, we are led to a conclusion that the observed wipe-out is apparently of a different origin.
A plausible explanation is that a second magnetic component that is unobservable in the present NQR experiment emerges below $T_h$.
In order to detect and characterize it, we next turn to complementary $^{13}$C NMR measurements.


\subsection{$^{13}$C Nuclear Magnetic Resonance}
%

Similarly to previous TF $\mu$SR and $^{14}$N NMR measurements, \cite{zorko2011unconventional} the $^{63,\,65}$Cu NQR can only follow the component that is disappearing with decreasing temperature below $T_h$, while the assumed second component is undetectable.
NMR measurements on $^{13}$C nuclei, which are positioned at the center of the NCN group (Fig.~\ref{fig0}), being the most distant from the electronic magnetism of Cu$^{2+}$ ions and possessing the weakest coupling among all the nuclei in CuNCN, can overcome this drawback.
Furthermore, contrary to the $^{63,\,65}$Cu and $^{14}$N nuclei, the $^{13}$C nuclei are spin-1/2 entities. 
Therefore, they do not couple to the EFG tensor and their NMR spectra are solely affected by the magnetism of CuNCN.
\begin{figure}[t]
\includegraphics[trim = 0mm 0mm 0mm 0mm, clip, width=1\linewidth]{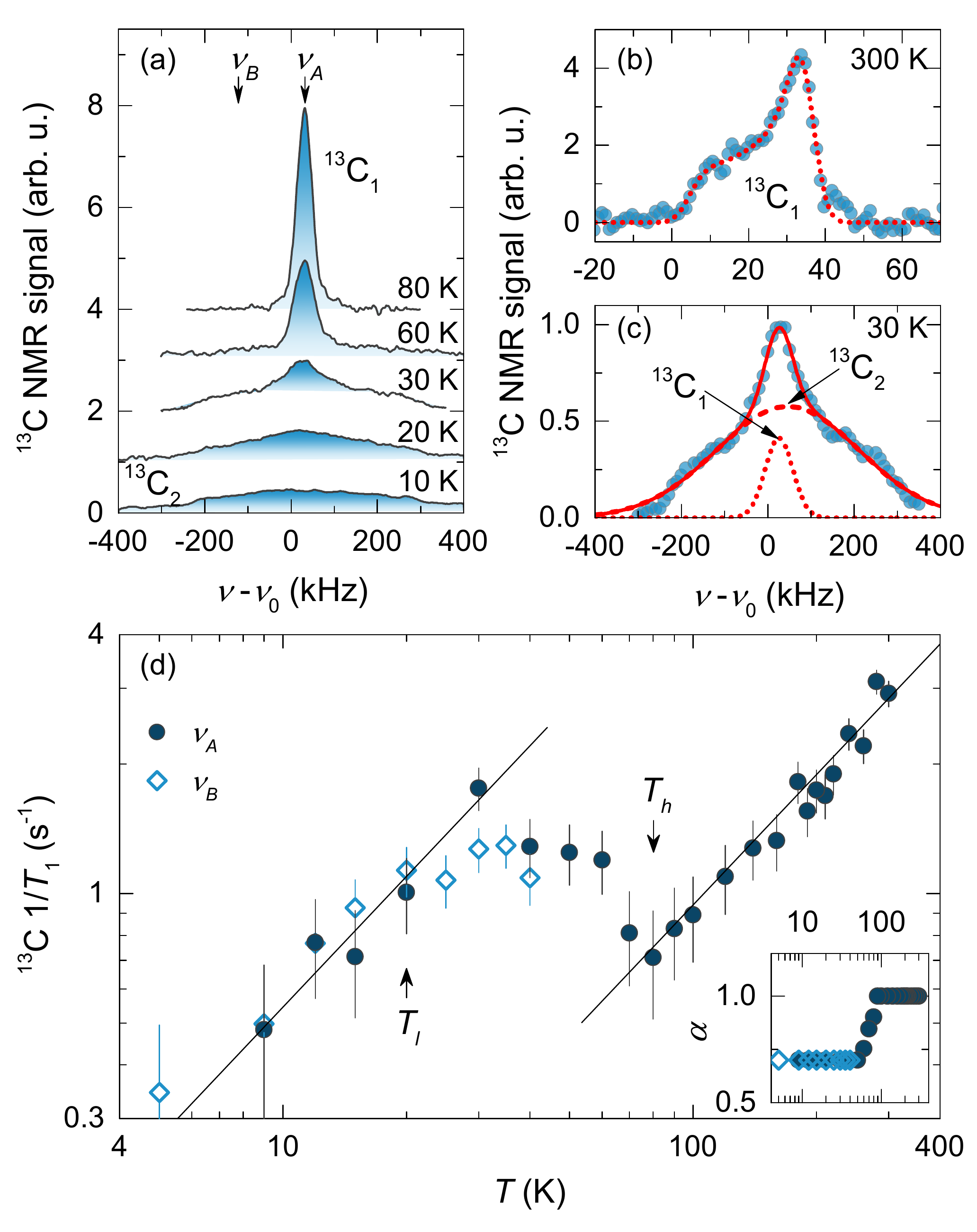}
\caption{(a) The $^{13}$C NMR spectra of CuNCN, with the two components ($^{13}$C$_1$ and $^{13}$C$_2$) appearing below $T_h=80$~K.
The spectra are shifted vertically for clarity and their intensity is normalized by the Boltzmann factor $1/T$.
(b) The powder simulation (dotted line) of the 300-K spectrum (circles) with axially-symmetric anisotropy. 
(c) The fit (solid line) of the 30-K spectrum (cycles) with two Gaussian components $^{13}$C$_1$ (dotted line) and $^{13}$C$_2$ (dashed line).
(d) The temperature dependence of the spin-lattice relaxation rate, measured at the positions $\nu_A$ and $\nu_B$, marked by arrows in panel (a).
The solid lines display a linear temperature dependence.
The arrows indicate the magnetically inhomogeneous region between $T_h$ and $T_l$.
Inset: the temperature dependence of the stretching exponent at the two frequencies.}
\label{fig2}
\end{figure}

At 300~K, the $^{13}$C NMR spectrum has an asymmetric powder line shape (Fig.~\ref{fig2}b), which is typical of a uniaxial local-field distribution that can either be due to an anisotropic hyperfine coupling or a chemical shift.
Simulation of the spectrum yields an isotropic local-field component of 240(5)~ppm, shifting the line from the $^{13}$C Larmor frequency $\nu_0$, and an anisotropic component of 200(5)~ppm, giving the line shape and line width. 
An additional isotropic Gaussian broadening with FWHM $^{13}\delta = 5.6$~kHz is needed, which we convolute with the powder spectrum to satisfactorily describe the NMR line shape.
With decreasing temperature the isotropic Gaussian broadening becomes dominant and completely overshadows the anisotropic line shape below $\sim$100~K, where the Gaussian FWHM reaches $^{13}\delta=25$~kHz. 
This pronounced magnetic broadening of the $^{13}$C NMR spectrum is reminiscent of the broadening found in $^{14}$N NMR spectra, while the increase of the magnetic susceptibility in the same temperature range is much more moderate.\cite{zorko2011unconventional}
The isotropic magnetic broadening implies inhomogeneous local magnetic fields, most likely due to short-range magnetic correlations.

The shape of the $^{13}$C NMR spectrum changes drastically
below $T_h$ (Fig.~\ref{fig2}a), where a much broader component ($^{13}$C$_2$) accompanies the narrower high-temperature component ($^{13}$C$_1$).
The intensity of the $^{13}$C$_2$ component grows on the expense of the $^{13}$C$_1$ component, which is steadily disappearing when temperature is decreased (Fig.~\ref{fig2}a).
The $^{13}$C NMR spectra can be fit with two overlapping Gaussian components (Fig.~\ref{fig2}c) down to $T_l=20$~K, where the $^{13}$C$_1$ component completely disappears.
The broad component is approximately 4-times broader than the narrow one (see section \ref{muSR} for details on the quantitative analysis).
This gradual transformation of the $^{13}$C NMR spectrum  corroborates with the above-presented wipe-out of the $^{63}$Cu NQR and previous $\mu$SR results\cite{zorko2011unconventional} (Fig.~\ref{fig1}b). 
Importantly, it also provides the first direct insight into the low-temperature phase, which other measurements could not detect.
The static local field at the $^{13}$C nuclear site, estimated from the FWHM $^{13}\delta_2=370$~kHz of the broad $^{13}$C$_2$ component at 5~K, is $\pi\,^{13}\delta_2/^{13}\gamma=17$~mT.

The two $^{13}$C NMR components are further inspected on the basis of their spin-lattice relaxation rates $1/T_1$ (Fig.~\ref{fig2}d).
Above $T_h=80$~K, i.e., in the magnetically homogeneous phase, we find a single-exponential relaxation at the frequency $\nu_A$, corresponding to the maximum of the narrow $^{13}$C$_1$ NMR component.
The relaxation rate increases linearly with temperature.
Below $T_h$, a stretched-exponential relaxation (stretching exponent $\alpha<1$; see the inset in Fig.~\ref{fig2}d) is observed, while $1/T_1$ changes its trend and starts increasing with decreasing temperature.
Since the amplitude of the broad $^{13}$C$_2$ component is very small at the $\nu_A$ position down to $\sim$50~K (Fig.~\ref{fig2}a), we attribute this inverse trend to the narrow $^{13}$C$_1$ NMR component.
Indeed, $T_1$ analysis where the spectrum-integration window around the $\nu_A$ position is reduced in steps from the original 80 down to 4~kHz does not yield any significant change of the $T_1$ values in this temperature range.
Below $\sim$50~K, the amplitudes of the two components at the $\nu_A$ position become comparable, therefore, the extracted relaxation rate is an average of both components.
We note that varying the integration window at these temperatures changes the $1/T_1$ value within less than 15\%, with the maximal variation occurring at 30~K. This implies that the relaxation rates of the two NMR components are similar in magnitude.
The observed changes in the stretching exponent $\alpha$ (inset in Fig.~\ref{fig2}d) demonstrate broadening of the distribution of relaxation rates at $\nu_A$ below $T_h$, something regularly encountered in inhomogeneous magnetic phases.
The same effect was previously observed by $^{14}$N NMR.\cite{zorko2011unconventional}

To avoid the overlap of the two components, we additionally performed $^{13}$C NMR $1/T_1$ measurements at the $\nu_B$ position, where the contribution of the narrow $^{13}$C$_1$ component is negligible at all temperatures (Figs.~\ref{fig2}a, \ref{fig2}c).
Therefore, we attribute this relaxation solely to the broad $^{13}$C$_2$ component.
The measured relaxation rate $1/T_1$ at $\nu_B$ increases with temperature in a quasi-linear fashion, and a stretched-exponential relaxation ($\alpha=0.7$) is found, similar to the $\nu_A$ position.
We note that the relaxation at $\nu_B$ could not be measured reliably at temperatures above 40~K due to weak signal.

The $^{13}$C $1/T_1$ rate normalized by the factor $(^{13}\gamma)^2$, which is proportional to the magnitude of the square of local fluctuating magnetic fields, is about three orders of magnitude below the corresponding value for $^{14}$N at room temperature.\cite{zorko2011unconventional}
This can be rationalized with a significantly reduced spin density at the $^{13}$C site compared to the $^{14}$N site.
Such a reduction of the spin density, despite the fact that the exchange interaction bridged by the NCN$^{2- }$ group should be extremely large,\cite{tsirlin2010uniform, tchougreeff2013low} is expected based on local-density approximation (LDA) calculations.
These predict that the highest occupied molecular orbital (HOMO) of NCN$^{2- }$, overlapping with the Cu$^{2+}$ orbital, is composed mainly of the two nitrogen orbitals due to $\pi$ bonding, while the contribution of the carbon orbitals is minor.\cite{tsirlin2010uniform}  
Secondly, filtering effects are possibly important already at room temperature, as the predicted dominant antiferromagnetic exchange interaction along the $c$ crystallographic axis exceeds room temperature. \cite{tsirlin2010uniform,tchougreeff2013low}
The position of the $^{13}$C nucleus on the NCN$^{2-}$ bond with respect to the surrounding copper moments is more symmetric than the position of the $^{14}$N nucleus (Fig.~\ref{fig0}). 
As a result, the $^{13}$C nucleus filters out antiferromagnetic correlations along the $c$ axis mediated by the isotropic part of the transferred hyperfine coupling, while the $^{14}$N nucleus does not.
The significantly reduced hyperfine coupling on the $^{13}$C site compared to other nuclear sites in CuNCN also explains why both magnetic components that appear below $T_h$ can be observed by $^{13}$C NMR.
In contrast, only the dynamical high-temperature component can be detected by $^{63}$Cu NQR and $^{14}$N NMR,\cite{zorko2011unconventional} while the low-temperature component possessing larger frozen fields is inaccessible by these two types of nuclei.

\subsection{Muon Spin Rotation}\label{muSR}
To further characterize the two magnetic components appearing below $T_h$ we turn to $\mu$SR.
Previous zero-field (ZF) $\mu$SR experiments on CuNCN demonstrated that the internal field $B_{int}$ in the low-temperature magnetic phase is frozen and broadly distributed.\cite{zorko2011unconventional,tsirlin2012hidden}
The distribution width (18~mT at 5~K) was found to be similar to the average field magnitude.  
Due to the inhomogeneous magnetism below $T_h$, it was essentially impossible to fit the ZF data unequivocally between $T_h$ and $T_l$.
The TF measurements, on the other hand, have been performed so far only in a weak applied field $B_{TF}=2\,{\rm mT}\ll B_{int}$.
In this case, only the signal of the dynamical high-temperature component could be traced below $T_h$ and was gradually lost on decreasing temperature (Fig.~\ref{fig1}b), while the $\mu$SR signal corresponding to the magnetically frozen component was undetectable.\cite{zorko2011unconventional}
Further insight into the inhomogeneous magnetism, complementary to the previous $\mu$SR investigations and present $^{63}$Cu NQR and $^{13}$C NMR, can be gained from TF measurements in a strong external field ($B_{int}\ll B_{TF}$), where $B_{int}$ represents only a small perturbation to $B_{TF}$.
\begin{figure}[t]
\includegraphics[trim = 0mm 0mm 0mm 0mm, clip, width=1\linewidth]{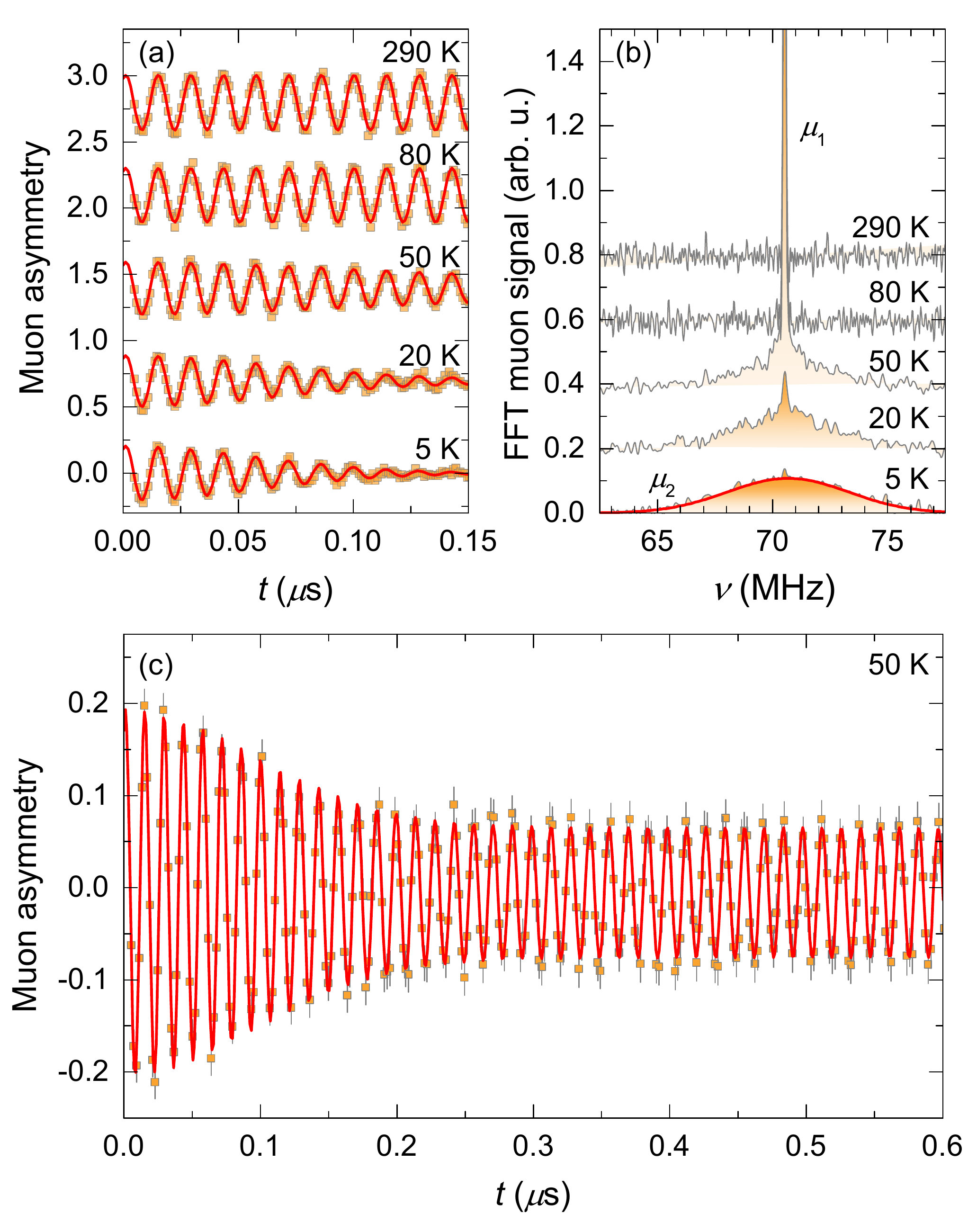}
\caption{(a) Oscillation of muon asymmetry in the transverse magnetic field of 520~mT and (b) the corresponding Fourier transforms.
The curves are shifted vertically for clarity.
(c) The $\mu$SR signal at 50~K, demonstrating the presence of two oscillating  component with significantly different damping rates.
The solid lines in (a) and (c) correspond to fits with two damped cosine contributions [Eq.~(\ref{eq1})], while in (b) the line represents a Gaussian fit.}
\label{fig3}
\end{figure}

The muon asymmetry $A$ measured in the applied field $B_{TF}=520$~mT shows a pronounced evolution with temperature (Fig.~\ref{fig3}a).
At 290~K, a single-component slowly relaxing oscillating signal is observed.
On the other hand, at temperatures below $T_h=80$~K, two oscillating components are clearly seen. 
The first component retains slow relaxation, while the second component relaxes much faster (Fig.~\ref{fig3}c). 
This can be further visualized by the Fourier transform analysis of the  oscillations (Fig.~\ref{fig3}b).
The analysis reveals a single-component ($\mu_1$) narrow spectrum above $T_h$ and an additional broad component ($\mu_2$) appearing below this temperature.
The intensity of the $\mu_1$ component decreases below $T_h$ with decreasing temperature and completely disappears below $T_l=20$~K.
These results are thus in complete agreement with the $^{13}$C NMR spectra also revealing the simultaneous presence of both the narrow and the broad component between $T_h$ and $T_l$. 
At 5~K, the $\mu$SR FFT spectrum is Gaussian (Fig.~\ref{fig3}b). 
Its FWHM of $\delta_\mu=5.3$~MHz corresponds to a static-field-distribution width of $\pi \delta_\mu/\gamma_\mu =20$~mT, which is in perfect agreement with previous ZF $\mu$SR results.\cite{zorko2011unconventional}
We note that internal fields in the range 10--100~mT are typically detected by muons in frozen magnetic insulators with spin-1/2 entities.\cite{yaouanc2011muon} 
Therefore, the frozen moment in the broad $\mu_2$ component must represent a significant part of a full Bohr magneton.
On the contrary, the width of the narrow $\mu_1$ component is about 25-times smaller (see below), thus once more revealing its predominantly dynamical nature.

\begin{figure}[b]
\includegraphics[trim = 0mm 28mm 0mm 0mm, clip, width=1\linewidth]{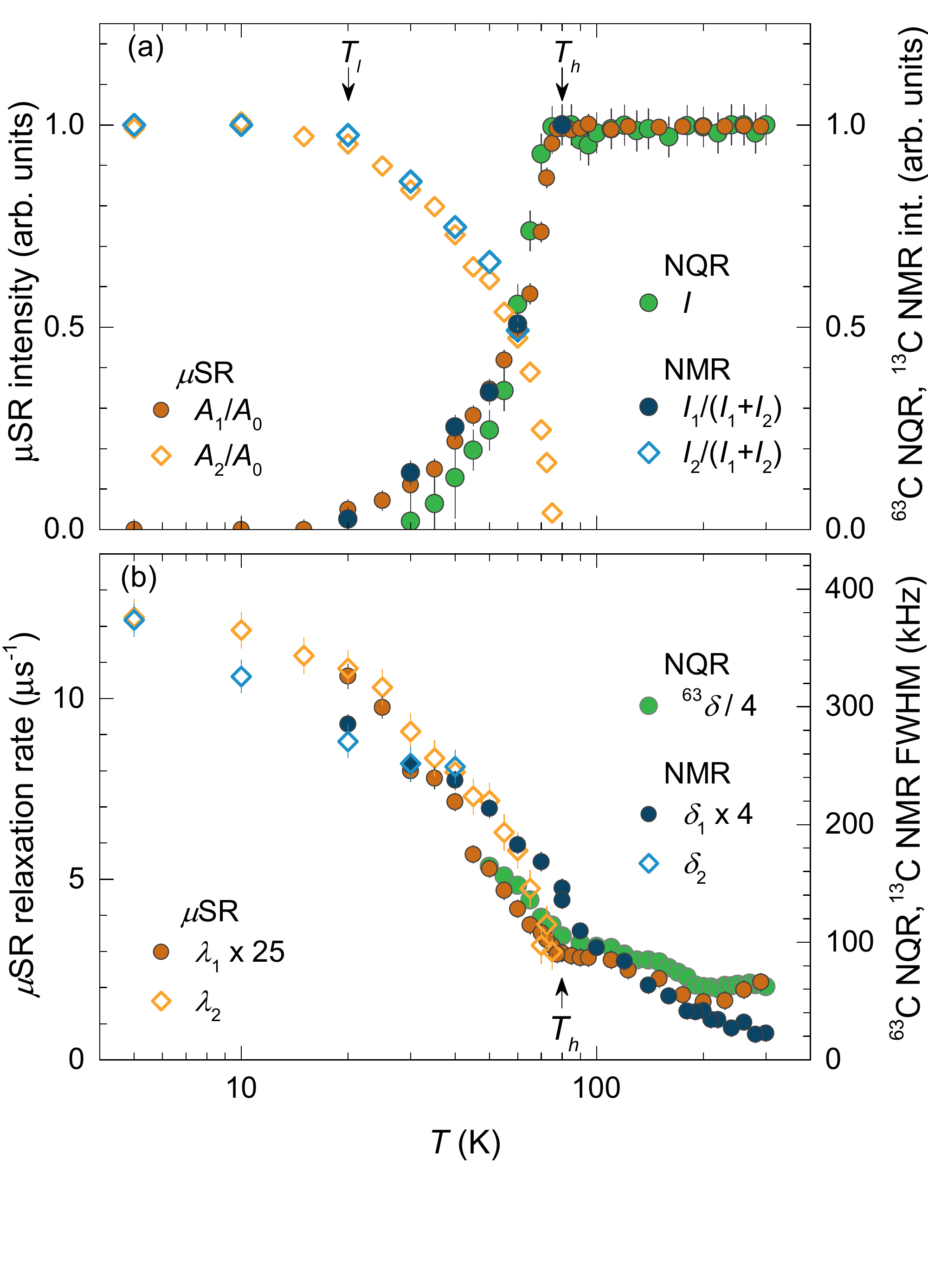}
\caption{(a) The temperature dependence of the normalized intensity of the narrow $\mu_1$ and the broad $\mu_2$ $\mu$SR component (defined as $A_i(T)/A_0$, where $A_0$ corresponds to full asymmetry) compared to the relative intensities $I_i/(I_1+I_2)$ of the narrow $^{13}$C$_1$ and broad $^{13}$C$_2$ NMR component and the $^{63}$C NQR intensity.
(b) Comparison of the $\mu$SR relaxation rates, the $^{13}$C NMR line widths of the narrow (1) and broad (2) components and the $^{63}$C NQR line width.
The arrows indicate specific temperatures $T_h$ and $T_l$.}
\label{fig4}
\end{figure}

A more quantitative insight is obtained from fitting the muon data in the time domain (Fig.~\ref{fig3}a).
The muon asymmetry can be modeled with a single damped cosine component above $T_h$ and below $T_l$, while two such components with different relative amplitudes $A_i$, frequencies $\nu_i$, and relaxation rates $\lambda_i$ are needed in the intermediate temperature regime, where
\begin{equation}
\label{eq1}
A(t) =A_1\cos \left(2\pi\nu_1 t \right){\rm e}^{-(\lambda_1 t)^2}
+A_2\cos \left(2\pi\nu_2 t \right){\rm e}^{-(\lambda_2 t)^2}.
\end{equation}
The two amplitudes are found to sum to the full asymmetry of $A_0=0.206$ at all temperatures. 
Their temperature dependence, shown in Fig.~\ref{fig4}a, quantifies -- in terms of volume fractions -- the gradual transition from the dynamical $\mu_1$ to the frozen $\mu_2$ component between $T_h$ and $T_l$.
Identical results (Fig.~\ref{fig4}a) are obtained from fitting the complementary $^{13}$C NMR spectra with two Gaussian contributions (Fig.~\ref{fig2}c) between $T_l$ and 60~K, where the fits become unreliable due to weak intensity of the broad component.
The disappearance of the high-temperature component (1) and its broadening with decreasing temperatures in both experiments  (Fig.~\ref{fig4}b) also corresponds nicely to the behavior of the $^{63}$Cu NQR signal. 

The temperature evolution of the $\mu$SR relaxation rates $\lambda_i$ and $^{13}$C line widths $^{13}\delta_i$ of the two components is displayed in Fig.~\ref{fig4}b. 
We note that both $\lambda_i$'s exceed the longitudinal muon relaxation rate\cite{tsirlin2012hidden} and both $^{13}\delta_i$'s exceed the $^{13}$C spin-lattice relaxation rates by at least an order of magnitude.
$\lambda_i$'s and $^{13}\delta_i$'s in general contain a contribution due to a static-field distributions and a contribution due to dynamical fields fluctuating at the Larmor frequency, while the longitudinal muon relaxation and the nuclear spin-lattice relaxation are solely due to dynamical fields.\cite{abragam1961principles}
Therefore, the static-local-field contribution to $\lambda_i$ and $^{13}\delta_i$ must dominate.
The static-field distributions increase sizably already in the dynamical high-temperature magnetic component (1) when decreasing the temperature from room temperature towards $T_h=80$~K (Fig.~\ref{fig4}b), implying the presence of short-range correlations.
Below $T_h$ the increasing trend of $\lambda_1$ and $^{13}\delta_1$ with decreasing temperature becomes even more pronounced.
Interestingly, this increase is correlated with the increasing relaxation/width of the other, frozen component (2) that emerges below $T_h$.


\section{Discussion}
Let us first compare our results to previous magnetic studies of CuNCN.
Tsirlin {\it et al.}~dubbed the low-temperature phase of CuNCN a "hidden magnetic order".\cite{tsirlin2012hidden} 
We have indeed found complementary evidence of frozen magnetism at low temperatures in this study.
However, in addition to the fact that magnetic Bragg peaks are absent,\cite{xiang2009theoretical} there are several experimental findings that speak strongly against long-range magnetic order.
First, we emphasize that the width of the distribution of the internal static fields is extremely large, i.e., of the order of the average field magnitude.
This property, initially revealed by ZF $\mu$SR experiments\cite{zorko2011unconventional,tsirlin2012hidden} is here confirmed by the broad Gaussian TF $\mu$SR and $^{13}$C NMR spectra at low temperatures. 
In contrast, box-shaped spectra would be observed for a powder sample in the case of a well-defined internal field.
Secondly, we stress that the $^{13}$C spin-lattice relaxation rate at low temperatures shows a linear temperature dependence, which also does not agree with ordinary 3D antiferromagnetically ordered states.
Namely, for the usually dominant Raman-magnon-scattering process one generally finds $1/T_1\propto T^3$ for temperatures much above the gap in the spin-wave spectrum. \cite{beeman1968nuclear}
If the temperature is lower than the gap or higher-order magnon scattering terms are dominant the temperature dependence of $1/T_1$ will be even steeper.\cite{beeman1968nuclear}

These observations show that the low-temperature magnetic state of CuNCN is a frozen, highly magnetically disordered state.
In terms of static magnetic fields, this state is closer to a spin-glass-like state than to a long-range antiferromagnetic order.
However, the canonical spin-glass picture does not apply, since the onset of spin freezing at $T_h=80$~K seems to be independent of the magnetic field, as both the zero-field NQR and the NMR experiments performed in 9.4~T show essentially the same results.
This field independence and the lack of any zero-field-cooled/field-cooled magnetic irreversibility\cite{zorko2011unconventional} differentiate CuNCN from other more conventional disordered frozen spin systems.\cite{binder1986spin}

In another proposal, a competing scenario to a frozen magnetic state was presented.\cite{zorko2011unconventional} 
This scenario of a spin-liquid instability at $T_h=80$~K was proposed based on (i) the fact that the exchange coupling was predicted to be extremely large,\cite{liu2008characterization, tsirlin2010uniform} potentially allowing for a correlated magnetic state already at temperatures as high as room temperature, (ii) the spin-only magnetic susceptibility deduced from ESR was found to exhibit a sudden decrease at $T_h$,\cite{zorko2011unconventional}, (iii) even polarized neutron diffraction experiments failed to detect any magnetic Bragg peaks down to the lowest temperatures,\cite{xiang2009theoretical} and (iv) the spin liquids were possibly fragile to external perturbations, such as implanted muons in the $\mu$SR experiment and high magnetic fields applied in the NMR experiments.
The present experiments eliminate the uncertainties of the point (iv), unambiguously reveal frozen magnetic fields, and thus leave much less room for any interpretation based of a transition between a gapless high-temperature spin liquid and a pseudo-gapped low-temperature spin liquid.

All complementary experimental techniques employed in our study, even the zero-field NQR, show the same general features.
The data is in fact entirely consistent with an inhomogeneous state with two magnetic components below $T_h$, the essentially dynamical nature of one of the components and the presence of large static magnetic fields in the other component.
Such a two-component picture also explains the decrease of the ESR susceptibility\cite{zorko2011unconventional} below $T_h$. 
An interesting finding of the two experiments that are able to detect both magnetic components between $T_h$ and $T_l$ ($^{13}$C NMR and TF $\mu$SR) is that internal fields emerge also in the dynamical phase and that the broadening of the dynamical $^{13}$C$_1$ component $^{13}\delta_1$ and the relaxation of the dynamical $\mu_1$ component $\lambda_1$ scale with the broadening $^{13}\delta_2$ and relaxation $\lambda_2$ of the corresponding frozen components (Fig.~\ref{fig4}b).
This indicates that the dynamical component and the frozen component are intertwined rather then being phase segregated.
Furthermore, the scaling suggests that the small static fields observed in the dynamical component actually originate from the magnetically frozen component, implying a microscopic mixture of both components. 
We find that the scaling ratios $\lambda_2/\lambda_1\sim 25$ and $^{13}\delta_2/^{13}\delta_1\sim 4$ differ substantially (Fig.~\ref{fig4}b.
This is most likely due to the fact that the value of $^{13}\delta_2$ is relatively decreased due to a symmetric position of the $^{13}$C nucleus on the NCN$^{2-}$ bond with respect to the surrounding frozen copper moments (Fig.~\ref{fig0}). 
For the isotropic part of the hyperfine coupling this symmetry causes filtering-out of antiferromagnetic correlations in both $a$ and $c$ crystallographic directions, along which the exchange couplings are predicted to be dominant.\cite{tsirlin2010uniform,tchougreeff2013low} 
This does not apply for the usually much less symmetric position of the muon stopping site.\cite{yaouanc2011muon}
On the other hand, the ratio $^{63,65}\delta_2/^{63,65}\delta_1$ of the Cu NQR line widths should be enhanced compared to the $\lambda_2/\lambda_1\sim 25$ ratio.
The reason is that the $^{63,65}\delta_1$ line width of the detectable dynamical component originates from a transferred hyperfine coupling with 
neighboring static Cu$^{2+}$ electronic moments (Fig.~\ref{fig0}), while the contribution of dynamical on-site moments is reduced due to exchange narrowing. \cite{abragam1961principles} On the other hand, for the undetectable frozen component the $^{63,65}\delta_2$ line width is determined by the hyperfine coupling to the on-site static moments.
The on-site hyperfine coupling is usually orders of magnitude larger than the  transferred coupling. 
Therefore, the frozen $^{63,65}$Cu NQR component should be much broader than the dynamical component, which explains why we were unable to detect it. 

Similar magnetic inhomogeneities to the one observed in CuNCN between $T_h=80$~K and $T_l=20$~K have been related in the literature\cite{stewart2004phase, zheng2006coexisting, ling2017striped, nakajima2012microscopic, zorko2014frustration, zorko2015magnetic, nilsen2015complex, pregelj2015spin, pregelj2016exchange} to geometrical frustration of the underlying spin lattice leading to a degenerate ground-state manifold.
The coexistence of different magnetic components on a microscopic scale is then a natural way to release frustration.
Although the exact spin model of CuNCN is at present still a subject of debate,\cite{tsirlin2010uniform, tchougreeff2013low} the observation of magnetic inhomogeneity in the broad temperature range between $T_h$ and $T_l$ speaks in favor of a highly frustrated spin model, rather than any model where geometrical frustration is negligible.

The gradual onset of the frozen magnetic phase below $T_h=80$~K is reminiscent of disorder-broadened first-order phase transitions,\cite{manekar2001first, kumar2006relating} regularly encountered in the formation of metastable magnetic states with glassy characteristics, e.g., such as found in colossal magnetoressistive manganites.\cite{tokura1996competing,dagotto2005complexity}
However, the latter phenomenon is generally field dependent, shows thermal hysteresis, and relies on the presence of quenched disorder or accommodation strain.\cite{dagotto2005complexity, ahn2004strain, wu2006magnetic}
As already emphasized above, there is no apparent field dependence in spin freezing of CuNCN below $T_h$, at least for fields up to 9.4~T.
We also observed no thermal hysteresis in any of our experiments.
Furthermore, the fraction of  magnetic impurities in our sample was estimated to be as low as 0.04\%.\cite{zorko2011unconventional}
Therefore, the scenario of the disorder-broadened first-order phase transition that would result in an "undercooled" dynamical state in CuNCN below $T_h$ seems  unlikely.

On the other hand, we note that a recent high-resolution synchrotron study of CuNCN revealed anisotropic broadening of Bragg peaks, especially those with reflection indexes related to the crystallographic $b$ axis.\cite{tsirlin2012hidden} 
This suggests the presence of  microstructural irregularities, which,
surprisingly, are anticorrelated with the chemical inhomogeneity, as the anisotropic broadening is most pronounced in samples with the lowest amount of impurities and almost perfect stoichiometry.\cite{tsirlin2012hidden}
In this respect, CuNCN resembles the case of $\alpha$-NaMnO$_2$. 
There, a similar kind of anisotropic broadening of Bragg reflections was shown to originate from near-degenerate crystal structures where geometrical frustration of the spin lattice led to a magnetostructurally inhomogeneous phase-separated ground state.\cite{zorko2014frustration,zorko2015magnetic} 
Although only one stable crystallographic phase has been experimentally found in CuNCN so far, it is reassuring to note that two polymorphs exist in the case of a sister HgNCN compound,\cite{liu2002synthesis} where their energy difference is rather small.\cite{liu2003experimental}
Alternatively, extremely low-frequency flexural modes of CuNCN along the $b$ axis\cite{tchougreeff2017atomic} could also explain the anisotropic broadening of synchrotron Bragg peaks.\cite{tsirlin2012hidden} 
These or similar modes can also be involved in the formation of the inhomogeneous magnetic state, as observed in our experiments on a time scale longer than $\sim$0.1~$\mu$s, if the magnetoelastic coupling is substantial.

Finally, we highlight a signature of a structural change that is found in CuNCN in the Cu NQR experiment.
At $T^{*}=200$~K, a clear peak is observed in the spin-spin relaxation rate $1/T_2$ (inset in Fig.~\ref{fig1}d), which is accompanied by a notable line-width increase below the same temperature (Fig.~\ref{fig1}d). 
As no magnetic anomaly has been observed so far at 200~K in any experiment, we attribute this behavior to a subtle structural effect, like freezing-out of a particular lattice excitation. 
In CuNCN, flexural phonon modes and libration modes are limited to rather low frequencies/energies\cite{tchougreeff2017atomic} and may thus fall into the time-window of the $1/T_2$ measurements.

\section{Conclusions}
The combination of complementary local-probe techniques of $^{63,65}$Cu NQR, $^{13}$C NMR and TF $\mu$SR employed in this study has unveiled a remarkably complex magnetic state of CuNCN.
We have firmly established that the magnetic state below $T_h=80$~K is intrinsically inhomogeneous, as the same kind of behavior is observed by all three experimental techniques, including NQR, which presents no perturbation to the physical system whatsoever.
On decreasing temperature below $T_h$ towards $T_l=20$~K, the high-temperature dynamical component continuously transforms into the essentially disordered low-temperature frozen component, as evidenced by both the two-component $^{13}$C NMR spectra (Fig.~\ref{fig2}c) and the two-component $\mu$SR signal (Fig.~\ref{fig3}b).
Importantly, we find that the line width/damping of the dynamical component detected by $^{13}$C NMR/$\mu$SR scales with the magnetically frozen component.
This experimental finding demonstrates a mutual magnetic coupling between the two components and  
eliminates the possibility of phase segregation in favor of the two components coexisting on a microscopic scale.
Further in-depth investigations are needed to ultimately unveil the corresponding microscopic mechanism of the intriguing magnetic behavior of CuNCN.

\acknowledgments{The financial support of the Slovenian Research Agency under the program No.~P1-0125 and project No.~N1-0052 is acknowledged. AT acknowledges financial support of the Russian Foundation for Basic Research under project No.~18-29-04051.}
%
%
%
%
%
\end{document}